\documentclass[useAMS,usenatbib,usegraphicx]{mn2e}

\usepackage{graphicx}
\usepackage[latin1]{inputenc}
\usepackage{color}
\usepackage{times}
\usepackage{natbib}
\usepackage{setspace}
\newif\ifAMStwofonts
\AMStwofontstrue
\definecolor{red}{rgb}{1,0.,0.}

\newcommand{\munich}{{\sc l-galaxies }}
\newcommand{\lcdm}{$\Lambda$CDM }
\newcommand{\fr}{$f(R)$-Gravity }
\newcommand{\msun}{{\rm M}_\odot}

\def\lesssim{\lower.5ex\hbox{$\; \buildrel < \over \sim \;$}}
\def\gtrsim{\lower.5ex\hbox{$\; \buildrel > \over \sim \;$}}
\title[Galaxy formation and coupled dark energy] {Semi-Analytic Galaxy Formation
  in coupled Dark Energy Cosmologies}

\author[Fontanot et al.]{
  \parbox[t]{\textwidth}{Fabio Fontanot$^1$\thanks{E-mail:
      fontanot@oats.inaf.it}, Marco Baldi$^{2,3,4}$, Volker
    Springel$^{5,6}$ and Davide Bianchi$^7$}
    \vspace*{8pt}\\
    $^1$ INAF - Astronomical Observatory of Trieste, via G.B. Tiepolo 11, I-34143 Trieste, Italy \\
    $^2$ Dipartimento di Fisica e Astronomia, Alma Mater Studiorum Universit`a di Bologna, viale Berti Pichat, 6/2, I-40127 Bologna, Italy\\
    $^3$ INAF - Osservatorio Astronomico di Bologna, via Ranzani 1, I-40127 Bologna, Italy \\
    $^4$ INFN - Sezione di Bologna, viale Berti Pichat 6/2, I-40127 Bologna, Italy \\
    $^5$ HITS-Heidelberger Institut f\"ur Theoretische Studien, Schloss-Wolfsbrunnenweg 35, 69118 Heidelberg, Germany\\
    $^6$ Zentrum f\"ur Astronomie der Universit\"at Heidelberg, ARI, M\"onchhofstrasse 12-14, 69120 Heidelberg, Germany \\
    $^7$ Institute of Cosmology and Gravitation, University of Portsmouth, Portsmouth PO1 3FX\\
}

\begin{document}
\date{Accepted ... Received ...}

\maketitle

\begin{abstract} 
Among the possible alternatives to the standard cosmological model
(\lcdm), coupled Dark Energy models postulate that Dark Energy (DE),
seen as a dynamical scalar field, may interact with Dark Matter (DM),
giving rise to a ``fifth-force'', felt by DM particles only. In this
paper, we study the impact of these cosmologies on the statistical
properties of galaxy populations by combining high-resolution
numerical simulations with semi-analytic models (SAM) of galaxy
formation and evolution. New features have been implemented in the
reference SAM in order to have it run self-consistently and calibrated
on these cosmological simulations. They include an appropriate
modification of the mass temperature relation and of the baryon
fraction in DM haloes, due to the different virial scalings and to the
gravitational bias, respectively. Our results show that the
predictions of our coupled-DE SAM do not differ significantly from
theoretical predictions obtained with standard SAMs applied to a
reference \lcdm simulation, implying that the statistical properties
of galaxies provide only a weak probe for these alternative
cosmological models. On the other hand, we show that both galaxy bias
and the galaxy pairwise velocity distribution are sensitive to coupled
DE models: this implies that these probes might be successfully
applied to disentangle among quintessence, \fr and coupled DE models.
\end{abstract}

\begin{keywords}
  galaxies: formation - galaxies: evolution - galaxies:fundamental
  properties
\end{keywords}

\section{Introduction}\label{sec:intro}
Dark Energy (DE) represents a critical unknown for the concordance
cosmological model of our Universe, which emerged as the result of a
decade long effort in the determination of the key cosmological
parameters \citep[see e.g][]{Planck_cosmpar}. The easiest description
for this mysterious contributor to the present energy density, which
accounts for $\sim 70$ per cent, is a classical cosmological constant
$\Lambda$, i.e. an homogeneous and static energy density filling the
whole Universe at all epochs. This simple model (\lcdm hereafter),
while indeed able to explain the vast majority of the observed
properties of the Universe, bears a number of theoretical problems
\citep[see e.g.][for a review]{Weinberg89}, mostly due to the level of
``fine-tuning'' required to accommodate for the small value of
$\Lambda$ at the present epoch. A number of alternative DE models have
thus been proposed in the literature trying to explain the origin of
the accelerated expansion: these range from scalar field theories
(i.e. quintessence), to modifications of the equation of general
relativity \citep[see e.g.][and references therein for a comprehensive
  review]{Amendola13}. In order to provide the observational
constraints needed to disentangle between those different scenarios,
wide galaxy surveys are currently under advanced planning, like the
Euclid mission \citep{Laureijs11}, which relies on a combination of
weak lensing measurements (based on precision imaging) and clustering
analysis (from slitless spectroscopy), which will allow to study, at
the same time, both the evolution of the equation of state of DE and
the growth function.

Since galaxies provide the privileged tracers of cosmic evolution, it
is crucial, for the success of these missions, to understand the
interplay between the physical processes responsible for galaxy
formation and evolution and the assembly of the large-scale
structure. The former issue has been explored, for a number of
alternative cosmologies, with the help of numerical methods that
follow the non-linear evolution of virialized structures (i.e. high
resolution $N$-body simulations, see e.g. \citealt{GrossiSpringel09,
  Baldi12, Puchwein13} and references herein). In the baryonic sector,
semi-analytic models (SAMs) employ simplified analytic prescriptions
to model the relevant processes acting on the baryonic gas (and their
interplay), and to study the evolution of galaxy components as a
function of their physical properties, redshift and environment. This
approach has been shown to correctly reproduce a number of
observational data, and the predictions of different models are
consistent in many cases \citep{Fontanot09b, Fontanot12a}, even if
some tension with the data still remains \citep[see
  e.g.][]{McCarthy07, BoylanKolchin12, Weinmann12}. The analytic
prescriptions embedded into SAMs are physically grounded and
observationally motivated but involve numerous parameters, which are
constrained by requiring the model to reproduce a well defined sample
of (typically) low-z observations. This approach thus harbours a
significant level of degeneracies \citep[see e.g][]{Henriques09},
which are also related to the fact that different authors adopt
different approximations for key physical mechanisms. In the context
of future space missions, it is therefore of fundamental importance to
characterize the impact of alternative DE models on the predicted
properties of galaxy populations, in order to devise observational
tests that can safely distinguish different cosmological models.

Our group was the first to study the implications of non-standard
cosmological scenarios on the relevant properties of galaxies in a
large cosmological volume, and to fully quantify their impact on the
statistical properties of galaxy populations as predicted by SAMs. We
focus on the amplitude of the expected modifications in the galaxy
stellar mass function, on the cosmic star formation rate and on the
2-point correlation function, but we also consider higher order
statistics like galaxy bias and the galaxy pairwise velocity
distribution. For the cosmologies we tested so far, we concluded that
galaxy properties alone are usually inefficient to constrain
cosmological models beyond \lcdm~, but whenever they are combined with
suitable information on the underlying Dark Matter (DM) distribution,
it is possible to devise statistical tests able to disentangle these
alternative cosmological scenarios from the standard model {\it and}
among themselves. This is the fourth paper on our series: we already
considered Early Dark Energy \citep[][hereafter Paper I]{Fontanot12c},
$f(R)$-Gravity \citep[][hereafter Paper II]{Fontanot13b} and massive
neutrino cosmologies \citep[][hereafter Paper III]{Fontanot15}. In
this work, we consider a new class of cosmologies, namely the
so-called {\it coupled DE} models \citep[cDE hereafter, see
  e.g.][]{Wetterich95,Amendola00}, which are based on the dynamical
evolution of a classical scalar field $\phi$ that plays the role of
DE, and interacts directly with the Cold Dark Matter (CDM) fluid. As a
consequence of the exchange of energy and momentum during cosmic
evolution, the DE scalar field thus mediates a ``fifth-force'' between
Dark Matter particles, leading to a modification of the gravitational
growth process (both at linear and non-linear scales) and to a
different evolution of cosmic large-scale structure. N-body
simulations of cDE cosmologies have been performed by various groups
in the last decade \citep[see e.g.][]{Maccio04, Baldi10, LiBarrow11,
  Carlesi14} and represent now a robust tool to investigate DE
interactions in the non-linear regime. The particular imprints of this
class of models have been studied in detail by \citet{Baldi12} using a
suite of $N$-body and hydrodynamical simulations (the {\sc CoDECS}
project\footnote{The simulations data of the {\sc CoDECS} project are
  publicly available at www.marcobaldi.it/CoDECS}), focusing on the
degeneracy of DE coupling with other cosmological parameters (like the
normalization of the matter power spectrum $\sigma_8$) and on the
gravitational bias, i.e. the offset between the density fluctuations
in CDM and in the baryonic components (leading to different baryon
fraction at cluster scales).

This paper is organized as follows. In Section~\ref{sec:models}, we
introduce the cosmological numerical simulations and semi-analytic
models we use in our analysis. We present the predicted galaxy
properties and compare them among different cosmologies in
Section~\ref{sec:results}. Finally, we discuss our conclusions in
Section~\ref{sec:final}.

\section{Models}\label{sec:models}
\begin{table}
  \caption{Cosmological parameters for our simulations. The columns
    contain from left to right: the normalization $A$ and the exponent
    $\alpha$ of the potential, the strength of the coupling $\beta$,
    $\sigma_8$ and the equation of state parameter $w$ at $z=0$.}
  \label{tab:runs}
  \renewcommand{\footnoterule}{}
  \centering
  \begin{tabular}{ccccccccc}
    \hline
     & A & $\alpha$ & $\beta$ & $\sigma_8(z=0)$ & $w(z=0)$ \\
    \hline
    \lcdm &  ---   & ---  &  ---  & 0.809 & -1.0   \\
    EXP003  & 0.0218 & 0.08 &  0.15 & 0.967 & -0.992 \\
    SUGRA003  & 0.0202 & 2.15 & -0.15 & 0.806 & -0.901 \\
    \hline
  \end{tabular}
\end{table}

\subsection{Coupled Dark Energy cosmologies}

In this work, we consider a set of flat cosmological models including
CDM, baryons, radiation and a DE scalar field $\phi$. Among the range
of models included in the {\sc CoDECS} project we focus on two
different choices for the scalar field potential. We first consider an
exponential potential \citep[][the EXP003 model in
  \citealt{Baldi12}]{Wetterich88}

\begin{equation}
V(\phi) = A \phi e^{-\alpha \phi},
\end{equation}

\noindent
which is characterized by stable scaling solutions for the scalar
field independently from initial conditions. In particular, in cDE
scenarios, such a potential provides a transient early DE solution and
a late time accelerated attractor \citep[see e.g.][]{Amendola04}.  We
also consider a SUGRA potential \citep[][the SUGRA003 model in
  \citealt{Baldi12}]{BraxMartin99}

\begin{equation}
V(\phi) = A \phi^{-\alpha} e^{-\phi^2/2}
\end{equation}

\noindent
which is typical of supersymmetric theories of gravity and implies a
``bounce'' of the DE equation of state $w$ at the cosmological
``barrier'' $w=-1$ \citep[see e.g.][]{Baldi12a}. This feature has
relevant implications for the expected number density evolution of DM
haloes, as well as for the evolution of the cosmological Hubble
function and growth factor \citep[see][for more details]{Baldi12}.

The evolution of the main cosmological components is described by a
set of dynamical equations including the interaction between the
scalar field and CDM particles (i.e. the right-hand side of
Equations~\ref{eq:cde1} and~\ref{eq:cde2}):

\begin{equation}\label{eq:cde1}
\ddot{\phi} + 3 H \dot{\phi} + \frac{dV}{d\phi} = \sqrt{\frac{16 \pi G}{3}} \beta(\phi) \rho_{\rm c}
\end{equation}

\begin{equation}\label{eq:cde2}
\dot{\rho_{\rm c}} + 3 H \rho_{\rm c} = - \sqrt{\frac{16 \pi G}{3}} \beta(\phi) \rho_{\rm c} \dot{\phi}
\end{equation}

\noindent
where $\rho_c$ represents the density of CDM particles and an overdot
denotes the (cosmic) time derivative. In the cDE models considered in
this work, the coupling function $\beta(\phi)$, which controls the
interaction strength\footnote{In particular, the sign of $\dot{\phi}
  \beta(\phi)$ is related to the energy-momentum flow between the two
  components, such as negative (positive) values of this quantity
  correspond to a transfer from DE to CDM (from CDM to DE) and to an
  increase (decrease) of the DM particle mass.}, is assumed to be
constant. As baryons and radiations are always uncoupled from $\phi$
their evolution follows the usual relations $\rho_{\rm b} \propto
a^{-3}$ and $\rho_{\rm r} \propto a^{-4}$, respectively.

At the level of linear perturbations, the interaction determines a
modification of the growth rate due to the presence of a fifth-force
with a strength $4\beta ^{2}/3$ times the standard gravitational
acceleration acting between CDM particles, and to an additional
velocity-dependent acceleration proportional to $\dot{\phi }\beta
$. These effective modifications of the standard gravitational
evolution affect also the non-linear dynamics of collapsed
structures. For a more detailed discussion on the linear and
non-linear properties of cDE cosmologies, we refer the reader to
\citet{Amendola04}, \citet{Baldi10} and \citet{Baldi11b}.

\subsection{Numerical simulations}
In this work, we take advantage of the results of the {\sc CoDECS}
numerical simulations \citep{Baldi12}, using a modified version of the
{\sc gadget} code \citep{Springel05}, designed to include the specific
physical processes arising in the cDE scenario \citep{Baldi10}. In
particular, we analyse the outcome of the {\sc H-CoDECS} set,
i.e. adiabatic hydrodynamical simulations on periodic boxes
$80\,h^{-1}{\rm Mpc}$ on a side, using $2 \times 512^3$ particles
(corresponding to a mass resolution of $2.39 \times 10^8\,h^{-1}{\rm
  M}_\odot$ for CDM and $4.78 \times 10^7\,h^{-1}{\rm M}_\odot$ for
baryons). An entropy-conserving formulation of smoothed particle
hydrodynamics \citep[SPH]{SpringelHernquist02} has been used to
estimate hydrodynamical forces acting on gas particles, and no
additional radiative processes (gas cooling, star formation,
feedbacks) have been included. Initial conditions for all runs were
generated using {\sc N-GENIC} by displacing particles from a
homogeneous {\it glass} distribution imposing the same amplitude of
the initial power spectrum at the last scattering surface, the same
phases and mode amplitudes, to ensure a similar realisation of the
large scale structure and to allow an object-by-object comparison. For
all simulations, a flat cosmological model has been assumed with $z=0$
cosmological parameters consistent with the 7th year results of the
{\it Wilkinson Microwave Anisotropy Probe} \citep[WMAP7]{WMAP7},
i.e. density parameters $\Omega_{\rm CDM}=0.226$, $\Omega_{\rm
  DE}=0.729$ and $\Omega_{\rm bar}=0.0451$ (for CDM, DE and baryons
respectively), Hubble parameter $h=0.703$ and Gaussian density
fluctuations with a scale-invariant primordial power spectrum with
spectral index $n=0.966$. The common normalisation of perturbations at
CMB and the different growth factors for cDE runs imply that the
EXP003 run has a different amplitude of density perturbation at every
redshift and a different $\sigma_8$ at $z=0$, as listed in
Table~\ref{tab:runs}. On the other hand, the SUGRA003 run has, by
construction, the same linear normalisation as \lcdm both at CMB and
at present, with the final results that its $\sigma_8$ value is very
similar to the \lcdm run.

For each run, 69 snapshots were stored\footnote{At variance with our
  previous work, the snapshot list is not the same as in the
  Millennium simulation; the {\sc CoDECS} redshift sampling was chosen
  mostly to allow the construction of full light-cones for weak
  lensing and CMB lensing purposes \citep[see e.g.][and references
    herein]{Giocoli15}}; the corresponding group catalogues were
generated using a Friend-of-Friend algorithm with a linking length of
0.2 (in mean particle separation units), and gravitationally bound
substructures have been identified using {\sc subfind}
\citep{Springel01} (only subhaloes that retain at least 20 particles
after the gravitational unbinding procedure were considered). We then
used the subhalo catalogues to define the merger tree histories as in
\citet{Springel05}.

\subsection{Semi-Analytic Models}
\begin{figure}
  \centerline{ \includegraphics[width=9cm]{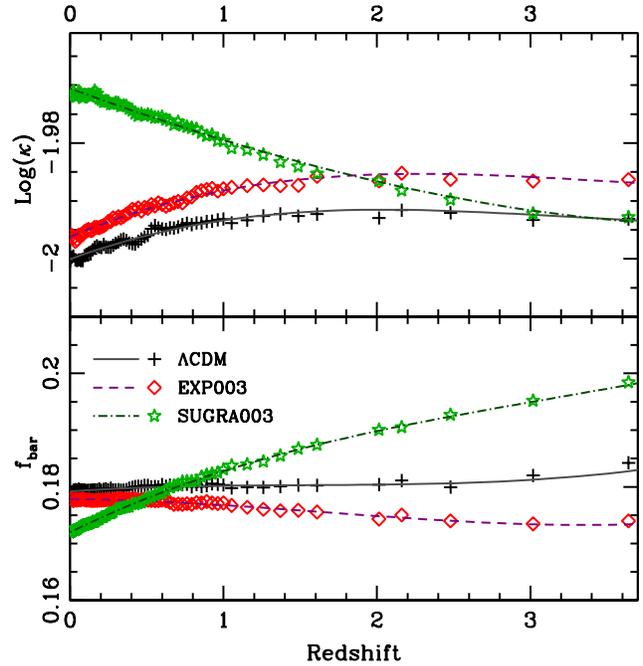} }
  \caption{Redshift evolution of the normalization of the virial
    scaling relations for DM haloes ($\kappa$, upper panel) and baryon
    fractions ($f_{\rm bar}$, lower panel). In each panel the black
    crosses, red diamonds and green stars refer to the \lcdm~, EXP003
    and SUGRA003 runs respectively. Solid, dashed and dot-dashed lines
    represent the 3-rd order polynomial best fits for each cosmology
    as indicated in the legends.}\label{fig:fits}
\end{figure}
\begin{figure*}
  \centerline{ \includegraphics[width=18cm]{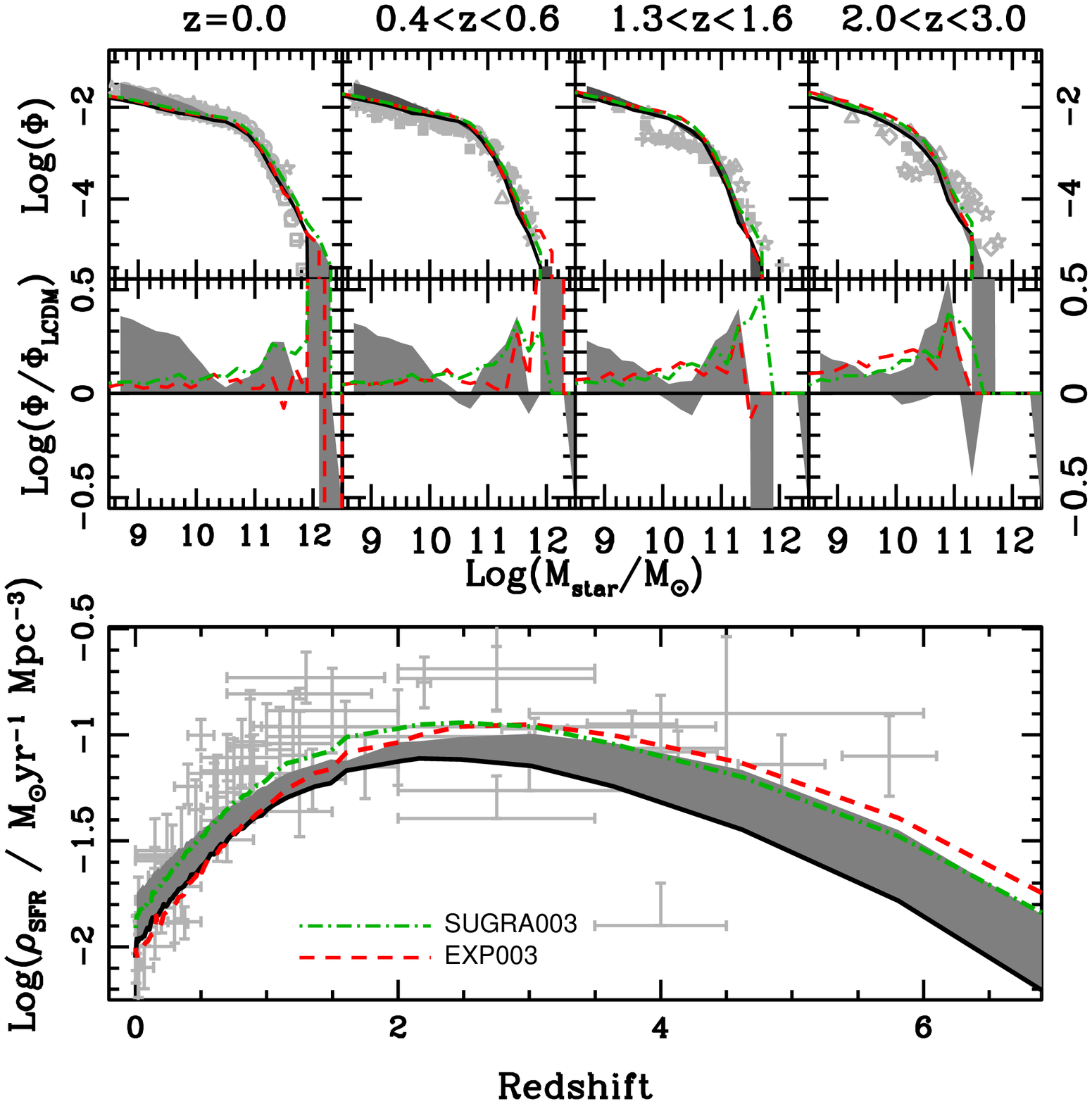} }
  \caption{Predicted galaxy properties for different coupled Dark
    Energy cosmological scenarios. {\it Upper panels:} redshift
    evolution of the stellar mass function (light grey points refer to
    the compilation from \citet{Fontanot09b}). The lower row shows the
    ratio between the mass function in a given cosmological model and
    the corresponding mass function from the \citet{Guo11} model in
    the \lcdm run. {\it Lower panel:} Cosmic star formation rate
    density (light grey points refer to the compilation from
    \citet{Hopkins04}). In each panel the solid black, dashed red and
    dot-dashed green lines refer to SAM predictions in \lcdm~, EXP003
    and SUGRA003 cosmologies respectively. Dark grey areas mark the
    distribution of the predictions from the \citet{Guo11},
    \citet{DeLuciaBlaizot07} and \citet{Croton06} SAMs applied to the
    same \lcdm run.}\label{fig:cde}
\end{figure*}
In this paper, we consider the same SAM suite we used in the previous
papers of the series. This includes three different versions of the
\munich semi-analytic model, namely those presented in
\citet{Croton06}, \citet{DeLuciaBlaizot07} and \citet{Guo11}. All
these models were run by construction on Millennium-like merger
trees\footnote{Thus avoiding any additional noise in the predictions
  due to different definitions of DM merger trees \citep{Knebe15}.}
as defined in the previous section and they represent a coherent set
of models\footnote{These three models mark the historical evolution of
  the code originally developed by \citet{Springel05}: from
  \citet{Croton06} to \citet{DeLuciaBlaizot07} the main changes
  involve the treatment of dynamical friction and merger times, the
  initial mass function (from Salpeter to Chabrier) and the dust
  modelling; while from \citet{DeLuciaBlaizot07} to \citet{Guo11} the
  modelling of supernovae feedback, the treatment of satellite galaxy
  evolution, tidal stripping and mergers were added.}, suitable to
study the intra-model variance due to different assumptions made in
the modelling of the key physical processes. The free parameters
usually associated with this approach have been calibrated (for a
\lcdm cosmological model), by comparing model predictions to a well
defined set of observational constraints (mainly at low-redshift). In
order to highlight the differences in galaxy properties due to the
different underlying cosmology, we choose not to recalibrate the
models, thus holding the role of other astrophysical processes
fixed. This implies that the models are optimally calibrated to
reproduce observations only for the \lcdm run, where we expect the
scatter in their predictions to be representative.

In the next section, we present the predictions obtained from a
modified version of the \citet{Guo11} model\footnote{For the sake of
  simplicity, in the following, we still refer to our modified code as
  the \citet{Guo11} model} run on the cDE boxes. The main changes with
respect to the \lcdm SAM version include the following new
features. First, the code handles a user-generated Hubble function
(tabulated in an external file) extracted from the corresponding
numerical realization of the cDE cosmology under analysis. In Paper
II, we showed that, for cosmologies introducing a fifth force, the
virial scaling relations deviate from the \lcdm expectations
\citep[see also][]{Arnold14}, i.e. the relation between total DM mass
inside a sphere with interior mean density 200 times the critical
density at a given redshift and the one-dimensional velocity
dispersion inside the same radius\footnote{\citet{Evrard08} showed
  that haloes in \lcdm cosmology are expected to follow the
  theoretical scaling $\sigma_{200} \propto [ h(z) M_{200}(z)
  ]^{1/3}$.} is modified. As in Paper II, we find that the actual
relations in cDE cosmologies (defined using $>10^{12} M_\odot$ DM
haloes) are offset versions of \lcdm ones (see e.g. Fig.~1 in Paper
II). Nonetheless, at variance with the constant shift for unscreened
haloes we found in the \fr case, the offset for cDE cosmologies is
redshift dependent, and smaller in amplitude (roughly corresponding to
a 1 percent variation at most). We show the different values for the
normalization $\kappa$ of the fitted virial relations in
Fig.~\ref{fig:fits} (upper panel). As in Paper II, we then apply a
correction to the virial scalings assumed in \munich~, corresponding
to the ratio between the virial scalings in the desired cosmology and
the corresponding virial scalings in the \lcdm run with the same
initial conditions. We then model the evolution of the offsets as a
3-rd order polynomial as a function of redshift.

A specific feature of cDE cosmologies is the different baryon fraction
($f_{\rm bar}$) \citep{Baldi10} in massive haloes, due to the
different forces felt by the baryons and DM particles. We estimate
$f_{\rm bar}$ for $>10^{12} M_\odot$ DM haloes in our simulations and
show the redshift evolution of this quantity in Fig.~\ref{fig:fits}
(bottom panel). Our results confirm the \citet{Baldi10} results; we
also stress that $f_{\rm bar}$ evolves as a function of redshift, and
it does not show any DM mass dependence at fixed cosmic epoch. Typical
differences with respect to the \lcdm run are smaller than 3 percent
at $z<2$. In \munich~, $f_{\rm bar}$ mainly regulates the infall of
pristine gas when DM haloes grow by accretion from the surrounding
field and it is usually treated as a redshift independent free
parameter. In our modified code, we still keep the baryon fraction as
a free parameter, but we require it to scale, at a given redshift, as
the ratio between the corresponding $f_{\rm bar}$ in the cDE and \lcdm
boxes. Also in this case, we model the baryon fraction evolution in
different cosmologies as a 3-rd order polynomial in redshift. It is
worth stressing, that the $f_{\rm bar}$ shown in Fig.~\ref{fig:fits}
refers to the ratio between the mass in baryons and the DM mass,
yielding a definition closer to the quantities actually used in
\munich~. We also consider the alternative definition of $f_{\rm bar}$
as the ratio between the baryonic mass and the total mass in the halo:
for all cosmological models, the different definition mainly changes
the normalization of $f_{\rm bar}(z)$, but not its
evolution. Therefore, as our modifications to \munich involve only
fractional quantities, our results are insensitive to the $f_{\rm
  bar}$ definition employed.

It is also worth stressing that these two features we included in the
SAM have a limited (but not negligible) impact on model predictions,
due to the small offsets with respect to the \lcdm realization and
that most of the differences we will discuss in the following are
triggered by differences in merger trees and cosmic growth history of
the large scale structure realized in the cosmological volumes under
analysis.

\section{Results \& Discussion}\label{sec:results}
\begin{figure}
  \centerline{ 
    \includegraphics[width=9cm]{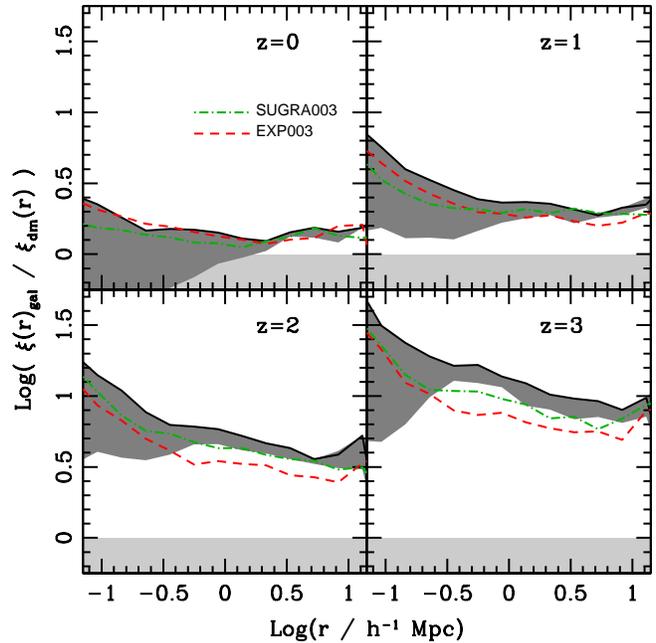} }
  \caption{Redshift evolution of galaxy bias. In each panel, only
    model galaxies with $M_\star > 10^9 \msun$ have been
    considered. Line types, colours and shaded areas have the same
    meaning as in Figure~\ref{fig:cde}.}\label{fig:bias}
\end{figure}
\begin{figure*}
  \centerline{ \includegraphics[width=18cm]{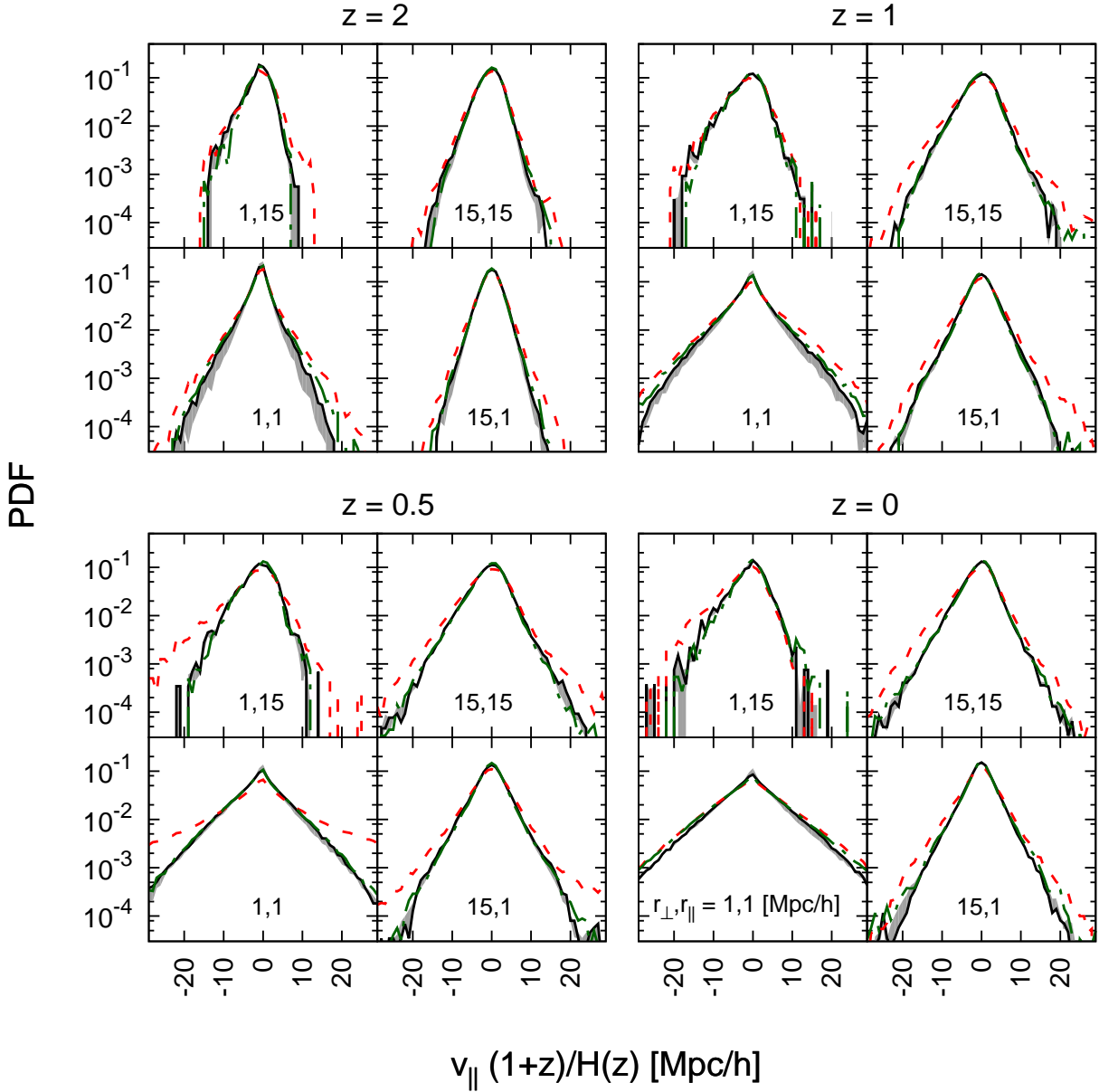} }
\caption{Pairwise galaxy velocity distribution along the line of sight
  for the cDE cosmologies considered in this paper at four different
  redshifts. Different cosmological models are marked by different
  linetypes, shading and colours as in Figure~\ref{fig:cde}. The upper
  left caption in each panel states the different combination for
  values of the galaxy separation ($r_\perp$,$r_\parallel$),
  perpendicular and parallel to the line of sight, respectively; in
  all panels, the conformal Hubble function $\mathcal{H}=aH$ has been
  used to rescale velocities to comoving
  distances.}\label{fig:pairwise}
\end{figure*}
In Figure~\ref{fig:cde}, we present a selection of the statistical
properties of $M_\star>10^9 M_\odot$ galaxies, as predicted by the
different SAMs. In the upper and lower panels, we show the redshift
evolution of the galaxy stellar mass function and cosmic star
formation rate, respectively. The model predictions have been
convolved with a lognormal error distribution with amplitude 0.25
(0.3) for stellar masses (star formation rates) to account for the
typical observational error in the estimate of these physical
quantities \citep{Fontanot09b}. Stellar masses and star formation
rates in the \citet{Croton06} model have been converted from Salpeter
to Chabrier IMF applying a constant shift of 0.25 and 0.176 dex,
respectively. In all panels, solid black lines refer to the
predictions of the \citet{Guo11} model for the \lcdm simulation. We
also consider the predictions of the other two SAMs applied to the
same simulation to estimate the scatter in the predictions of
different SAMs applied to the same cosmological box (shaded areas): we
stress again that the main source of this scatter lies in the
different treatment of the key physical mechanisms driving galaxy
evolution in the three models. In all panels, we also show the
predictions for the modified \citet{Guo11} model in the EXP003 and
SUGRA003 runs, with red dashed and green dot-dashed lines,
respectively. As far as galaxy properties are concerned, the
deviations of the model predictions in these alternative cosmological
scenarios with respect to the reference \lcdm run are quite small and
comparable to the intra-model variance at fixed \lcdm cosmology. This
provides an {\it a-posteriori} justification for our choice not to
recalibrate the model parameters, as the properties of the overall
galaxy populations are consistent among the different cosmological
runs. There is a slight tendency for the deviations to grow with
redshift, due to the different $\sigma_8$ value at $z=0$. At variance
with the \fr cosmology we studied in Paper II, we find no relevant
dependence of these deviations on stellar mass. In Paper II, we
interpret this effect as being due to the different virial scalings in
the \fr model, which directly affect the modeling of AGN
feedback. Here the deviations from the \lcdm virial scalings are
definitely smaller than in the \fr case, and the lack of a stellar
mass dependence in the ratio between the mass function in a given
cosmology and in the \lcdm run, clearly supports our conclusion that
this effect is negligible for the cDE cosmologies we consider in this
paper. All the differences seen in Fig.~\ref{fig:cde} are driven by
the different merger trees statistics associated with the different
cosmologies. Overall, the deviations from the \lcdm mass function
remain within 0.2 dex at most mass scales and redshifts.

In our previous work, we considered two additional cosmological tests,
namely galaxy bias and the pairwise velocity distribution, and we
discussed their efficiency in disentangling between \lcdm and other
cosmological scenarios. The discriminating power of these observables
is mainly driven by the combination of galaxy populations statistics
with information on the distribution of the DM in the underlying large
scale structure. In Figure~\ref{fig:bias}, we show the galaxy bias
estimates for the cosmological models considered in this work based on
the ratio between the auto-correlation function of galaxies in real
space $\xi_{\rm gal}$ and the corresponding $\xi_{\rm cdm}$ computed
for a randomly selected subsample of 1 percent of the CDM particles in
the cosmological box. Only the EXP003 model shows a deviation from the
\lcdm model larger than the SAM variance in the standard cosmology: in
particular this holds at $z>1$ and for scales larger than $\sim$1
$h^{-1}{\rm Mpc}$. The different $\sigma_8$ in the EXP003 run
contributes to this deviation, but it cannot totally account for it at
scales smaller than $\sim$10 $h^{-1}{\rm Mpc}$ and for $z>1$, as
shown by \citet[their Figure~5]{Baldi12}. Also the SUGRA003 model
shows clear deviations from the corresponding \lcdm run, but those are
of the same order as the intra-SAM variance, thus limiting the
efficiency of this estimator in breaking the degeneracies between
different cosmologies.

Moreover, in Figure~\ref{fig:pairwise} we present the redshift
evolution for the pairwise galaxy velocity distribution along the line
of sight $\mathcal{P}(v_\parallel,r_\parallel,r_\perp)$, where we
consider fixed components of galaxy separation parallel
($r_\parallel$) and perpendicular ($r_\perp$) to the line of sight
\citep[see e.g.]{Scoccimarro04}. This quantity is a reliable indicator
of the assembly of the large scale structure, as it traces the
anisotropy of redshift-space correlation functions and it is strongly
sensitive to the abundance of massive haloes. In
Figure~\ref{fig:pairwise} we adopt the same reference separations we
choose in our previous work (i.e. 1 and 15 $h^{-1}{\rm Mpc}$), despite
the smaller box size of the {\sc H-CoDECS} runs (i.e. cubic boxes with
$80 \, h^{-1}{\rm Mpc}$ sides). Also in this plot, only galaxies with
$M_\star > 10^9 M_\odot$ have been considered; furthermore, the
velocities have been rescaled using the conformal Hubble function
$\mathcal{H}=aH$ in order for the distribution to represents the
statistical displacement of galaxy pairs from real to redshift
space. We assume the usual convention that the pairwise velocity is
positive when galaxies are receding and negative when they are
approaching. This analysis leads us to similar conclusions with
respect to Figure~\ref{fig:bias}: the EXP003 model clearly show a
different velocity distribution with respect to predictions in the
\lcdm run (due in part, but not completely, to the increase in
$\sigma_8$), while the SUGRA003 results are virtually
indistinguishable from those obtained for the standard cosmology.

\section{Conclusions}
\label{sec:final}

In this paper, we present an updated version of the \munich
semi-analytic model, designed to run self-consistently on the {\sc
  CoDECS} suite of numerical simulation for coupled Dark Energy
cosmologies \citep{Baldi12}. The main modifications with respect to
the \lcdm version of the code include: (a) an implementation of an
user-defined Hubble function; (b) modified DM halo virial scalings
(which impact the mass-temperature relation); (c) modified baryon
fractions accounting for the gravitational bias. Item (a) was already
introduced in Paper II; item (b) has been modified for this project by
allowing a redshift dependence for the virial scaling normalization
(with respect to the \lcdm expectations); item (c) has been introduced
in this work. All these new features of the model have been directly
calibrated on the cDE simulations under consideration, by comparison
with the corresponding \lcdm runs.

The new code represents a step forward with respect to previous
versions, designed to run on Early Dark Energy (Paper I) and \fr
cosmologies (Paper II), allowing us to present the first comprehensive
picture of the impact signature of non-standard coupled-DE cosmologies
on the statistical properties of galaxy populations. The weak
cosmological constraints coming from direct comparison with model
galaxy properties (alone) confirm the robustness of the SAM
predictions against small variations in the cosmological framework
\citep[see also][]{Wang08,Guo11}. On the other hand, the modification
of the cosmological framework we consider has a substantial impact on
the growth and assembly of the large scale structure: therefore,
combining predicted galaxy properties with information on the
distribution of the underlying total mass distribution, it becomes
possible to disentangle different cosmological scenarios. In
particular, we focus on standard tests like galaxy bias and the galaxy
pairwise velocity distribution, and we show that coupled DE models can
be distinguished from \lcdm runs using {\it both} probes, unlike
quintessence models (in Paper I we showed that only bias is a
sensitive probe) and \fr models (in Paper II we showed that only the
pairwise velocity distribution is a sensitive probe). However, our
results also show that these cosmological tests are sufficiently
sensitive only for a subset of coupled DE cosmologies, in particular
for the exponential potential run\footnote{It is worth stressing that
  the particular model tested in this paper (EXP003) has a relatively
  strong value of coupling, excluded at about $3\sigma $ C.L. by the
  most recent CMB constrains. However, such large coupling values
  might be still viable in the presence of a substantial (but still
  reasonable) contribution of massive neutrinos to the cosmic
  budget.}, while the run including a supersymmetric potential
corresponds to a much weaker (although coherent) signal. Moreover, in
Paper III we showed that the likely existence of a massive neutrino
background implies deviations from a pure \lcdm run which go on in the
opposite direction with respect to these results, i.e. an increased
galaxy bias and a narrower pairwise distribution. Therefore, the
inclusion of such a background would have the net effect of reducing
the cosmological signal coming from either coupled DE or \fr models
(as seen i.e. in \citealt{Baldi14}).

Overall, the results we presented in our series of papers are of
particular relevance for the planning and exploitation of future wide
area galaxy surveys (like Euclid, \citealt{Laureijs11}), meant to
shade light on the true nature of DE. In fact, the wealth of data
coming from these efforts requires careful calibration and analysis in
order to provide a proper characterization of the large scale
structure evolution. Of course, a deep understanding of all the
systematic effects, either due to galaxy formation physics or the
cosmological parameters, is of critical importance, given the
exquisite precision required for disentangling the different
scenarios. Therefore, the construction of mock galaxy catalogues
covering the widest range of proposed DE theories represents a key
step. Our SAM suite, designed to run self-consistently on a variety of
these DE models, covers this need and provides a tool to test the
relative efficiency of cosmological probes. In this paper, as in our
previous work, we focus on scales suitable for galaxy studies
(i.e. stellar masses $10^9 M_\odot < M_{\rm star} < 10^{12} M_\odot$),
but using a moderate volume: we thus plan to apply our SAMs to larger
cosmological volumes in the future and use these runs to build
cosmological light cones \citep[see e.g.][]{Merson13} and mock galaxy
catalogues resembling, as close as possible the expected galaxy
properties in the different DE cosmologies.

\section*{Acknowledgements}
FF acknowledges financial support from the grants PRIN MIUR 2009 ``The
Intergalactic Medium as a probe of the growth of cosmic structures''
and PRIN INAF 2010 ``From the dawn of galaxy formation''. VS
acknowledges financial support from the Klaus Tschira Foundation and
the Deutsche Forschungsgemeinschaft through Transregio 33, ``The Dark
Universe''. During the development of this work MB has been partly
supported by the Marie Curie Intra European Fellowship ``SIDUN" within
the 7th Framework Programme of the European Commission.  The numerical
simulations presented in this work have been performed on the VIP and
on the Hydra clusters at the RZG supercomputing centre in Garching.

\bibliographystyle{mn2e}
\bibliography{fontanot}

\begin{thebibliography}{}

\bibitem[\protect\citeauthoryear{{Amendola}}{{Amendola}}{2000}]{Amendola00}
{Amendola} L.,  2000, \mnras, 312, 521

\bibitem[\protect\citeauthoryear{{Amendola}}{{Amendola}}{2004}]{Amendola04}
{Amendola} L.,  2004, \prd, 69, 103524

\bibitem[\protect\citeauthoryear{{Amendola}}{{Amendola}}{2013}]{Amendola13}
{Amendola} L. e.~a.,  2013, Living Reviews in Relativity, 16, 6

\bibitem[\protect\citeauthoryear{{Arnold}, {Puchwein} \& {Springel}}{{Arnold}
  et~al.}{2014}]{Arnold14}
{Arnold} C.,  {Puchwein} E.,    {Springel} V.,  2014, \mnras, 440, 833

\bibitem[\protect\citeauthoryear{{Baldi}}{{Baldi}}{2011}]{Baldi11b}
{Baldi} M.,  2011, \mnras, 414, 116

\bibitem[\protect\citeauthoryear{{Baldi}}{{Baldi}}{2012a}]{Baldi12a}
{Baldi} M.,  2012a, \mnras, 420, 430

\bibitem[\protect\citeauthoryear{{Baldi}}{{Baldi}}{2012b}]{Baldi12}
{Baldi} M.,  2012b, \mnras, 422, 1028

\bibitem[\protect\citeauthoryear{{Baldi}, {Pettorino}, {Robbers} \&
  {Springel}}{{Baldi} et~al.}{2010}]{Baldi10}
{Baldi} M.,  {Pettorino} V.,  {Robbers} G.,    {Springel} V.,  2010, \mnras,
  403, 1684

\bibitem[\protect\citeauthoryear{{Baldi}, {Villaescusa-Navarro}, {Viel},
  {Puchwein}, {Springel} \& {Moscardini}}{{Baldi} et~al.}{2014}]{Baldi14}
{Baldi} M.,  {Villaescusa-Navarro} F.,  {Viel} M.,  {Puchwein} E.,  {Springel}
  V.,    {Moscardini} L.,  2014, \mnras, 440, 75

\bibitem[\protect\citeauthoryear{{Boylan-Kolchin}, {Bullock} \&
  {Kaplinghat}}{{Boylan-Kolchin} et~al.}{2012}]{BoylanKolchin12}
{Boylan-Kolchin} M.,  {Bullock} J.~S.,    {Kaplinghat} M.,  2012, \mnras, 422,
  1203

\bibitem[\protect\citeauthoryear{{Brax} \& {Martin}}{{Brax} \&
  {Martin}}{1999}]{BraxMartin99}
{Brax} P.~H.,  {Martin} J.,  1999, Physics Letters B, 468, 40

\bibitem[\protect\citeauthoryear{{Carlesi}, {Knebe}, {Lewis} \& {Wales} S. an
  d~{Yepes}}{{Carlesi} et~al.}{2014}]{Carlesi14}
{Carlesi} E.,  {Knebe} A.,  {Lewis} G.~F.,    {Wales} S. an d~{Yepes} G.,
  2014, \mnras, 439, 2943

\bibitem[\protect\citeauthoryear{{Croton}, {Springel}, {White}, {De Lucia},
  {Frenk}, {Gao}, {Jenkins}, {Kauffmann}, {Navarro} \& {Yoshida}}{{Croton}
  et~al.}{2006}]{Croton06}
{Croton} D.~J.,  {Springel} V.,  {White} S.~D.~M.,  {De Lucia} G.,  {Frenk}
  C.~S.,  {Gao} L.,  {Jenkins} A.,  {Kauffmann} G.,  {Navarro} J.~F.,
  {Yoshida} N.,  2006, \mnras, 365, 11

\bibitem[\protect\citeauthoryear{{De Lucia} \& {Blaizot}}{{De Lucia} \&
  {Blaizot}}{2007}]{DeLuciaBlaizot07}
{De Lucia} G.,  {Blaizot} J.,  2007, \mnras, 375, 2

\bibitem[\protect\citeauthoryear{{Evrard}, {Bialek}, {Busha}, {White}, {Habib},
  {Heitmann}, {Warren}, {Rasia}, {Tormen}, {Moscardini}, {Power}, {Jenkins},
  {Gao}, {Frenk}, {Springel}, {White} \& {Diemand}}{{Evrard}
  et~al.}{2008}]{Evrard08}
{Evrard} A.~E.,  {Bialek} J.,  {Busha} M.,  {White} M.,  {Habib} S.,
  {Heitmann} K.,  {Warren} M.,  {Rasia} E.,  {Tormen} G.,  {Moscardini} L.,
  {Power} C.,  {Jenkins} A.~R.,  {Gao} L.,  {Frenk} C.~S.,  {Springel} V.,
  {White} S.~D.~M.,    {Diemand} J.,  2008, \apj, 672, 122

\bibitem[\protect\citeauthoryear{{Fontanot}, {Cristiani}, {Santini}, {Fontana},
  {Grazian} \& {Somerville}}{{Fontanot} et~al.}{2012}]{Fontanot12a}
{Fontanot} F.,  {Cristiani} S.,  {Santini} P.,  {Fontana} A.,  {Grazian} A.,
  {Somerville} R.~S.,  2012, \mnras, 421, 241

\bibitem[\protect\citeauthoryear{{Fontanot}, {De Lucia}, {Monaco}, {Somerville}
  \& {Santini}}{{Fontanot} et~al.}{2009}]{Fontanot09b}
{Fontanot} F.,  {De Lucia} G.,  {Monaco} P.,  {Somerville} R.~S.,    {Santini}
  P.,  2009, \mnras, 397, 1776

\bibitem[\protect\citeauthoryear{{Fontanot}, {Puchwein}, {Springel} \&
  {Bianchi}}{{Fontanot} et~al.}{2013}]{Fontanot13b}
{Fontanot} F.,  {Puchwein} E.,  {Springel} V.,    {Bianchi} D.,  2013, \mnras,
  436, 2672

\bibitem[\protect\citeauthoryear{{Fontanot}, {Springel}, {Angulo} \&
  {Henriques}}{{Fontanot} et~al.}{2012}]{Fontanot12c}
{Fontanot} F.,  {Springel} V.,  {Angulo} R.~E.,    {Henriques} B.,  2012,
  \mnras, 426, 2335

\bibitem[\protect\citeauthoryear{{Fontanot}, {Villaescusa-Navarro}, {Bianchi}
  \& {Viel}}{{Fontanot} et~al.}{2015}]{Fontanot15}
{Fontanot} F.,  {Villaescusa-Navarro} F.,  {Bianchi} D.,    {Viel} M.,  2015,
  \mnras, 447, 3361

\bibitem[\protect\citeauthoryear{{Giocoli}, {Metcalf}, {Baldi}, {Meneghetti},
  {Moscardini} \& {Petkova}}{{Giocoli} et~al.}{2015}]{Giocoli15}
{Giocoli} C.,  {Metcalf} R.~B.,  {Baldi} M.,  {Meneghetti} M.,  {Moscardini}
  L.,    {Petkova} M.,  2015, ArXiv e-prints (arXiv:1502.03442)

\bibitem[\protect\citeauthoryear{{Grossi} \& {Springel}}{{Grossi} \&
  {Springel}}{2009}]{GrossiSpringel09}
{Grossi} M.,  {Springel} V.,  2009, \mnras, 394, 1559

\bibitem[\protect\citeauthoryear{{Guo}, {White}, {Boylan-Kolchin}, {De Lucia},
  {Kauffmann}, {Lemson}, {Li}, {Springel} \& {Weinmann}}{{Guo}
  et~al.}{2011}]{Guo11}
{Guo} Q.,  {White} S.,  {Boylan-Kolchin} M.,  {De Lucia} G.,  {Kauffmann} G.,
  {Lemson} G.,  {Li} C.,  {Springel} V.,    {Weinmann} S.,  2011, \mnras, 413,
  101

\bibitem[\protect\citeauthoryear{{Henriques}, {Thomas}, {Oliver} \&
  {Roseboom}}{{Henriques} et~al.}{2009}]{Henriques09}
{Henriques} B.~M.~B.,  {Thomas} P.~A.,  {Oliver} S.,    {Roseboom} I.,  2009,
  \mnras, 396, 535

\bibitem[\protect\citeauthoryear{{Hopkins}}{{Hopkins}}{2004}]{Hopkins04}
{Hopkins} A.~M.,  2004, \apj, 615, 209

\bibitem[\protect\citeauthoryear{{Knebe} \& et al.}{{Knebe} \&
  et~al.}{2015}]{Knebe15}
{Knebe} A.,  et al. 2015, in preparation

\bibitem[\protect\citeauthoryear{{Komatsu}, {Smith}, {Dunkley}, {Bennett},
  {Gold} et~al.,}{{Komatsu} et~al.}{2011}]{WMAP7}
{Komatsu} E.,  {Smith} K.~M.,  {Dunkley} J.,  {Bennett} C.~L.,  {Gold} B.,
  et~al., 2011, \apjs, 192, 18

\bibitem[\protect\citeauthoryear{{Laureijs}, {Amiaux}, {Arduini},
  {Augu{\`e}res}, {Brinchmann}, {Cole}, {Cropper}, {Dabin}, {Duvet}, {Ealet} \&
  et al.}{{Laureijs} et~al.}{2011}]{Laureijs11}
{Laureijs} R.,  {Amiaux} J.,  {Arduini} S.,  {Augu{\`e}res} J.~.,  {Brinchmann}
  J.,  {Cole} R.,  {Cropper} M.,  {Dabin} C.,  {Duvet} L.,  {Ealet} A.,    et
  al. 2011, ArXiv e-prints (arXiv:1110.3193)

\bibitem[\protect\citeauthoryear{{Li} \& {Barrow}}{{Li} \&
  {Barrow}}{2011}]{LiBarrow11}
{Li} B.,  {Barrow} J.~D.,  2011, \prd, 83, 024007

\bibitem[\protect\citeauthoryear{{Macci{\`o}}, {Quercellini}, {Mainini},
  {Amendola} \& {Bonometto}}{{Macci{\`o}} et~al.}{2004}]{Maccio04}
{Macci{\`o}} A.~V.,  {Quercellini} C.,  {Mainini} R.,  {Amendola} L.,
  {Bonometto} S.~A.,  2004, \prd, 69, 123516

\bibitem[\protect\citeauthoryear{{McCarthy}, {Bower} \& {Balogh}}{{McCarthy}
  et~al.}{2007}]{McCarthy07}
{McCarthy} I.~G.,  {Bower} R.~G.,    {Balogh} M.~L.,  2007, \mnras, 377, 1457

\bibitem[\protect\citeauthoryear{{Merson}, {Baugh}, {Helly}, {Gonzalez-Perez},
  {Cole}, {Bielby}, {Norberg}, {Frenk}, {Benson}, {Bower}, {Lacey} \&
  {Lagos}}{{Merson} et~al.}{2013}]{Merson13}
{Merson} A.~I.,  {Baugh} C.~M.,  {Helly} J.~C.,  {Gonzalez-Perez} V.,  {Cole}
  S.,  {Bielby} R.,  {Norberg} P.,  {Frenk} C.~S.,  {Benson} A.~J.,  {Bower}
  R.~G.,  {Lacey} C.~G.,    {Lagos} C.~d.~P.,  2013, \mnras, 429, 556

\bibitem[\protect\citeauthoryear{{Planck Collaboration XVI}}{{Planck
  Collaboration XVI}}{2014}]{Planck_cosmpar}
{Planck Collaboration XVI} 2014, \aap, 571, A16

\bibitem[\protect\citeauthoryear{{Puchwein}, {Baldi} \& {Springel}}{{Puchwein}
  et~al.}{2013}]{Puchwein13}
{Puchwein} E.,  {Baldi} M.,    {Springel} V.,  2013, \mnras, 436, 348

\bibitem[\protect\citeauthoryear{{Scoccimarro}}{{Scoccimarro}}{2004}]{Scoccima%
rro04}
{Scoccimarro} R.,  2004, \prd, 70, 083007

\bibitem[\protect\citeauthoryear{{Springel} \& {Hernquist}}{{Springel} \&
  {Hernquist}}{2002}]{SpringelHernquist02}
{Springel} V.,  {Hernquist} L.,  2002, \mnras, 333, 649

\bibitem[\protect\citeauthoryear{{Springel}, {White}, {Jenkins}, {Frenk},
  {Yoshida}, {Gao}, {Navarro}, {Thacker}, {Croton}, {Helly}, {Peacock}, {Cole},
  {Thomas}, {Couchman}, {Evrard}, {Colberg} \& {Pearce}}{{Springel}
  et~al.}{2005}]{Springel05}
{Springel} V.,  {White} S.~D.~M.,  {Jenkins} A.,  {Frenk} C.~S.,  {Yoshida} N.,
   {Gao} L.,  {Navarro} J.,  {Thacker} R.,  {Croton} D.,  {Helly} J.,
  {Peacock} J.~A.,  {Cole} S.,  {Thomas} P.,  {Couchman} H.,  {Evrard} A.,
  {Colberg} J.,    {Pearce} F.,  2005, \nat, 435, 629

\bibitem[\protect\citeauthoryear{{Springel}, {White}, {Tormen} \&
  {Kauffmann}}{{Springel} et~al.}{2001}]{Springel01}
{Springel} V.,  {White} S.~D.~M.,  {Tormen} G.,    {Kauffmann} G.,  2001,
  \mnras, 328, 726

\bibitem[\protect\citeauthoryear{{Wang}, {De Lucia}, {Kitzbichler} \&
  {White}}{{Wang} et~al.}{2008}]{Wang08}
{Wang} J.,  {De Lucia} G.,  {Kitzbichler} M.~G.,    {White} S.~D.~M.,  2008,
  \mnras, 384, 1301

\bibitem[\protect\citeauthoryear{{Weinberg}}{{Weinberg}}{1989}]{Weinberg89}
{Weinberg} S.,  1989, Reviews of Modern Physics, 61, 1

\bibitem[\protect\citeauthoryear{{Weinmann}, {Pasquali}, {Oppenheimer},
  {Finlator}, {Mendel}, {Crain} \& {Macci{\`o}}}{{Weinmann}
  et~al.}{2012}]{Weinmann12}
{Weinmann} S.~M.,  {Pasquali} A.,  {Oppenheimer} B.~D.,  {Finlator} K.,
  {Mendel} J.~T.,  {Crain} R.~A.,    {Macci{\`o}} A.~V.,  2012, \mnras, 426,
  2797

\bibitem[\protect\citeauthoryear{{Wetterich}}{{Wetterich}}{1988}]{Wetterich88}
{Wetterich} C.,  1988, Nuclear Physics B, 302, 668

\bibitem[\protect\citeauthoryear{{Wetterich}}{{Wetterich}}{1995}]{Wetterich95}
{Wetterich} C.,  1995, \aap, 301, 321

\end{thebibliography}

\end{document}